# Heat flowing from cold to hot without external intervention by using a "thermal inductor"

A. Schilling[1*], X. Zhang[1], and O. Bossen[1]

[1]Department of Physics, University of Zürich, Winterthurerstrasse 190,
Zürich CH-8057, Switzerland

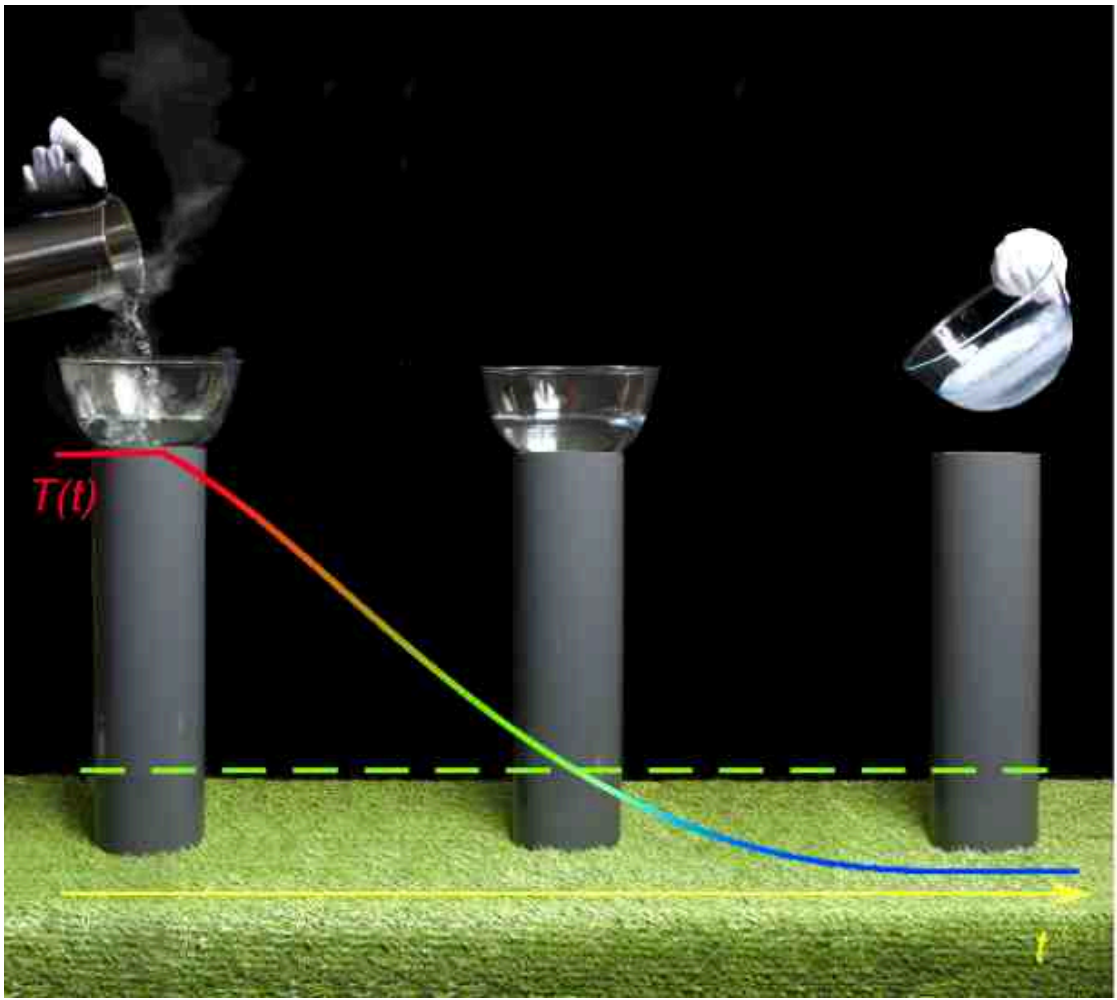

# Heat flowing from cold to hot without external intervention by using a "thermal inductor"

A. Schilling[1*], X. Zhang[1], and O. Bossen[1]

[1]Department of Physics, University of Zürich, Winterthurerstrasse 190,
Zürich CH-8057, Switzerland

*Corresponding author. Email: schilling@physik.uzh.ch

**The cooling of boiling water all the way down to freezing, by thermally connecting it to a thermal bath held at ambient temperature without external intervention, would be quite unexpected. We describe the equivalent of a "thermal inductor", composed of a Peltier element and an electric inductance, which can drive the temperature difference between two bodies to change sign by imposing inertia on the heat flowing between them, and enable continuing heat transfer from the chilling body to its warmer counterpart without the need of an external driving force. We demonstrate its operation in an experiment and show that the process can pass through a series of quasi-equilibrium states while fully complying with the second law of thermodynamics. This "thermal inductor" extends the analogy between electrical and thermal circuits and could serve, with further progress in thermoelectric materials, to cool hot materials well below ambient temperature without external energy supply or moving parts.**

**INTRODUCTION**
According to the rules of classical thermodynamics, the flow of heat between two thermally connected objects with different temperatures is determined by Fourier's law of heat conduction, stating that the rate of heat flow between these objects increases with growing temperature difference, and, more generally, by the second law of thermodynamics which requires that heat can flow by itself only from a warmer to a colder body. The majority of measures for an energy-efficient thermal management in everyday infrastructure are based on these laws, be it for the thermal insulation of buildings and heat accumulators, or for harvesting a maximum of mechanical work from a heat engine operating between two different temperatures. While the validity of these fundamental laws is undisputed, there have been exciting technical developments during the past few years which appear to be at odds with popular interpretations of these laws. In the simplest version of Fourier's law, for example, the rate of heat ($\dot{Q}$) flowing between a body at a temperature $T_b$ that is connected to an object with a different temperature $T_r$, is given by $\dot{Q} = k(T_b - T_r)$, where the thermal conductance $k$ of the connection is expected to be independent of the sign of the temperature difference $T_b - T_r$.

Nevertheless, a number of recent experiments demonstrated that a thermal rectification is possible to some extent, thereby opening up the way for a customized thermal management beyond the simple form of Fourier's law (*1*). The operation of such a "thermal diode" (i.e., an analogue to a diode rectifying electric current) has been demonstrated, for example, by the use of phononic devices (*2-6*), phase change materials (*7-10*), the application of quantum dots (*11*) and of engineered metallic hybrid devices at low temperatures (*12*), paving the way for even more sophisticated thermal circuits such as thermal transistors and logic gates (*6,13,14*).

The second law of thermodynamics, on the other hand, imposes strict limits on the efficiency of heat engines, cooling devices and heat pumps (*15*), and suggests a preferred direction of the flow of heat to reach a thermodynamic equilibrium (*16*). According to the latter interpretation, a hot body with temperature $T_b$ that is connected to a colder object at temperature $T_r$, approaches thermodynamic equilibrium with strictly $T_b > T_r$, and $T_b$ is expected to monotonically fall as a function of time *t* because heat is not supposed to flow by itself from a cold to a warmer body (Figs. 1A and 1C). Any undershooting or oscillatory behavior of $T_b$ with respect to $T_r$ (Figs. 1B and 1D) with a reversing direction of the heat flow and transferring heat from cold to hot, would normally be ascribed to an active intervention to remove heat with external work to be done, or to a violation of the second law of thermodynamics.

In this article, we describe a simple thermal connection containing a novel thermal element, namely, the equivalent of a "thermal inductor" that we had originally designed for precise heat-capacity measurements in an actively driven thermal circuit (*17*). We show that it can also act in a passive way without any external or internally hidden source of power, and is able to drive the temperature difference between two massive bodies to change sign by imposing a certain thermal inertia on the flow of heat. We demonstrate in an experiment that such a process can occur through a series of quasi-equilibrium states, and show that it still fully complies with the second law of thermodynamics in the sense that the entropy of the whole system monotonically increases over time, albeit heat is temporarily flowing from cold to hot.

**RESULTS AND DISCUSSION**

**Modeling of the oscillating thermal circuit**

The considered thermal connection consists of an ideal electrical inductor with inductance *L* and a Peltier element with Peltier coefficient $\Pi = \alpha T$, forming a closed electrical circuit (*17*) (Fig. 2A). Here, $\alpha$ stands for the combined Seebeck coefficient of the used thermoelectric materials, and *T* is the absolute temperature of the considered junction between these materials. When an electric current *I* is flowing through, heat *Q* is generated or absorbed at a rate $\dot{Q} = \Pi I = \alpha T I$, respectively, depending on the direction of the current. A body with heat capacity *C* and a thermal reservoir (or two finite bodies) are each in thermal contact with the opposite sides of the Peltier element providing a thermal link by its thermal conductance *k*. The process is described by Kirchhoff's voltage law in Eq. 1a containing the generated thermoelectric voltage $\alpha(T_b - T_r)$, and by the thermal balance equations Eqs. 1b and 1c for the rates of heat removed from (or added to) one body ($\dot{Q}_b$) and to (from) the other body or the thermal reservoir ($\dot{Q}_r$), respectively,

$$L\dot{I} + RI = \alpha(T_b - T_r), \quad (1a)$$
$$\dot{Q}_b = -\alpha T_b I + \tfrac{1}{2} RI^2 - k(T_b - T_r), \quad (1b)$$
$$\dot{Q}_r = +\alpha T_r I + \tfrac{1}{2} RI^2 + k(T_b - T_r), \quad (1c)$$

where $R$ is the internal resistance of the Peltier element and $\alpha$ is taken as a constant for simplicity. We also temporarily ignore parasitic effects due to other sources of electrical resistance or thermal transport through leads or convection, and the heat-capacity contribution of the Peltier element is thought to be absorbed in $C$. We count $\dot{Q} > 0$ for a heat input; the choice of the signs of $I$ and $\alpha$ in the Eqs. 1 turns out to be unimportant, however. The dissipated Joule heating power $RI^2$ is regarded to be equally distributed to both sides of the Peltier element. The individual contributions to the flow of heat in Eqs. 1b and 1c are visualized in Fig. 2B. The set of equations Eqs. 1, but without the inductive term $L\dot{I}$, is standard to describe the flow of heat and charge in a Peltier element (*18,19*).

To consider the situation for the actual experiment to be presented further below, where a finite heat capacity $C$ is connected to an infinite thermal reservoir as shown in Figs. 1A and 1B we combine $\dot{Q}_b = C\dot{T}_b$ and $T_r = const.$ with the Eqs. 1a and 1b and obtain a nonlinear differential equation for $I(t)$, namely

$$LC\ddot{I} + (RC + kL)\dot{I} + (kR + \alpha^2 T_r)I + \tfrac{1}{2}\alpha RI^2 + \alpha L I\dot{I} = 0. \quad (2)$$

This equation can be easily numerically solved with high accuracy. The time evolution $T_b(t)$ could then be obtained by inserting the corresponding solution $I(t)$ into Eq. 1a. For a systematic analysis we restrict ourselves to $\Delta_0 = T_b(0) - T_r \ll T_r$ with temperature independent $\alpha$, $k$, $R$ and $C$, valid within a sufficiently narrow temperature interval $\pm\Delta_0$ around $T_r$. Then, the last two terms of the differential equation Eq. 2 are negligible because with Eq. 1a, $\tfrac{1}{2}RI^2 + L I\dot{I} < RI^2 + L I\dot{I} = \alpha(T_b - T_r)I \ll \alpha T_r I$, and we end up with the equation for a damped harmonic oscillator,

$$LC\ddot{I} + (RC + kL)\dot{I} + (kR + \alpha^2 T_r)I \approx 0. \quad (3)$$

**Discussion of the oscillatory solutions**
It is very instructive to discuss at first the analytical solutions of this simplified equation. The initial conditions for $t = 0$, i.e., the time when the virgin thermal connection is inserted, are $I(0) = 0$ and $T_b(0) - T_r = \Delta_0$, thereby fixing $\dot{I}(0) = \alpha\Delta_0 / L$. The corresponding solution may show an over-damped or an oscillatory behavior, depending on the combination of the constant factors in the equation. To achieve a possible undershooting of $T_b(t)$ below $T_r$, we focus at first on oscillatory solutions of $I(t)$. The condition for their occurrence, $4\alpha^2 T_r > LC(k/C - R/L)^2$, can always be fulfilled for any value of $\alpha$, if $L$ is chosen as $L^* = RC/k$. While $R$ and $k$ are given by the characteristics of the Peltier element, $C$ is limited only by the heat capacity of the Peltier element but can otherwise be chosen at will. The solution of interest is $I(t) = I_0 \exp(-t/\tau)\sin(\omega t)$, with $\tau$ and $\omega$ matched to fulfill the differential equation, and $I_0 = \alpha \Delta_0 / (L\omega)$. The corresponding phase-shifted solution for the temperature is $T_b(t) - T_r = \Delta_0 \exp(-t/\tau)\cos(\omega t - \delta)/\cos\delta$ with $\tan\delta = (R - L/\tau)/L\omega$. We now seek the particular solution realizing the weakest possible damping of $I(t)$ within an oscillation cycle. This occurs for a maximum value of $\omega\tau$ where $L = L^*$, $\omega = \omega^* = \sqrt{\alpha^2 T_r k/(RC^2)}$ and $\tau = \tau^* = C/k$. Introducing the standard definition of the dimensionless figure of merit for a Peltier element at $T = T_r$ (20), $ZT_r = \alpha^2 T_r / kR = (\omega^* \tau^*)^2$ (hereafter abbreviated as $Z_T$), the first minimum of $T_b(t)$ is attained with these parameters at $t_{\min} = \beta \tau^*$ with $\beta = (\pi/2 + \arctan\sqrt{Z_T})/\sqrt{Z_T}$ for

$$T_{b,\min} = T_r - \Delta_0 \exp(-\beta)\sqrt{Z_T/(Z_T+1)} < T_r. \qquad (4)$$

If expressed by the dimensionless quantities $(T_b(t) - T_r)/\Delta_0$ and $t/\tau^*$, the time evolution and $(T_{b,\min} - T_r)/\Delta_0$, a measure for the maximum obtained cooling effect, only depend on $Z_T$ but are independent of the thermal load $C$ and the other parameters of the system. In Fig. 3A we show a series of corresponding $T_b(t)$ curves that we numerically obtained with Eq. 3 for different values of $Z_T$, expressed in these dimensionless units, with $T_b(0) - T_r = \Delta_0 = 80$ K and $T_r = 293$ K to mimic a realistic scenario. Despite the fairly large ratio $\Delta_0/T_r \approx 0.27$, the difference between the thus obtained values and the results calculated using the explicit solution of Eq. 3 with $\omega^* = \sqrt{Z_T}/\tau^*$ would be barely distinguishable in Fig. 3, however.

According to Eq. 4, the minimum temperature decreases with increasing $Z_T$, but is limited to $T_{b,\min} > T_r - \Delta_0$ for a finite value of $Z_T$ so that no catastrophic oscillation can occur. If the thermal connection were removed after the body has reached its minimum temperature, $T_b$ would stay at $T_{b,\min} < T_r$ under perfectly isolated conditions as sketched in Fig. 1B. Removing it in a state where $I = 0$ even leaves the thermal connection in its original, virgin state at $T_b = T_r - \Delta_0 \exp(-\pi/\sqrt{Z_T})$ but still with $T_b < T_r$ (inset of Fig. 3A). Any external work associated with the act of removing the thermal connection could be made infinitesimally small, e.g., by opening a nanometer-sized gap between the body and the thermal connection. If the connection is not removed at all, $T_b(t)$ exhibits a damped oscillation around $T_r$, approaching thermal equilibrium with eventually $T_b = T_r$. We note that the maximum possible cooling effect is not reached for the above parameters, but for $L_{opt} = \lambda L^*$ with $\lambda(Z_T) > 1$ (see

Supplementary Materials, section S1). The corresponding solutions for $I(t)$ are over-damped for $Z_T < 1/3$, but the temperature of the body still undershoots $T_r$ for all values of $Z_T > 0$.

In a closely related scenario, two finite bodies with identical heat capacities $2C$ and different initial temperatures $T_b(0)$ and $T_r(0)$ are thermally connected in the same way, and observed under completely isolated conditions (Fig. 1D). In the limit $\Delta_0 = T_b(0) - T_r(0) << \bar{T}$ with the mean initial temperature $\bar{T} = [T_b(0) + T_r(0)]/2$, we end up with the same simplified differential equation Eq. 3 for $I(t)$ but with $T_r$ replaced by $\bar{T}$ (see Supplementary Materials, section S2). In Fig. 3B, we show the resulting counter-oscillating temperatures $T_b(t)$ and $T_r(t)$ for $Z_T = 5$, together with the average temperature $T_{av} = [T_b(t) + T_r(t)]/2 \leq \bar{T}$, which is not constant but reaches local minima around the times when the two temperatures are equal.

**Oscillatory flow of heat**

Each time when $T_b - T_r$ changes its sign (this occurs for the first time as soon as $T_b$ drops below $T_r$, for $L = L^*$ after $t_0 = \pi/2\omega^*$, Figs. 1 and 3), heat is still continuously flowing from the cold to the warmer object (dark/purple arrows in Figs. 1 and 2B) until $|T_b - T_r|$ reaches its maximum, where the direction of the heat flow is reversed. The moving charge carriers drive virtually all of this heat directly away from the cold to the warm end, without being temporarily stored as energy of the magnetic field residing in the inductor. The maximum amount of magnetic energy, $\frac{1}{2}LI^2 < \frac{1}{2}LI_0^2 = \alpha^2 \Delta_0^2 / 2L^*\omega^{*2}$ is less than a fraction $\Delta_0/T_r << 1$ of the excess heat $\sim C\Delta_0$ that has been initially stored in the warmer body. In this sense, the electrical inductor acts, in interplay with the Peltier element, only as the driver of the temperature oscillation by imposing a certain thermal inertia that temporarily counteracts the flow of heat dictated by Fourier's law. Thus, we can interpret the role of the circuit as that of a "thermal inductor". In analogy to the self-inductance $L$ of an electrical inductor generating a voltage difference $\Delta V$ according to $L\dot{I} = -\Delta V$, we can even ascribe to the present circuit a thermal self-inductance $L_{th} = L/(\alpha^2 T_r)$ (*17*), obeying $L_{th}\dot{I}_{th} = L_{th}\ddot{Q} = -\Delta T$ (see Supplementary Materials, section S4).

From a refrigeration-engineering point of view, we can divide a full period of an oscillation cycle of $T_b(t)$ into four stages (see Fig. 2B). In the first quarter of a full period (*i*), the operation corresponds to that of a thermoelectric generator. In the second quarter (*ii*), the circuit acts as a thermoelectric cooler even for $T_b < T_r$, driven by the electric current that is still flowing in the original direction due to the action of the electric inductor. During the third quarter (*iii*), the electric current changes sign (thermoelectric generator), while heating persists in the fourth quarter (*iv*) even for $T_b > T_r$, and the device is then operating as a thermoelectric heater.

The order of magnitude of the rate of heat flowing between the objects can be chosen arbitrarily low by adjusting the thermal load $C$. In this way, the electronic oscillator circuit reaches the typically very large time scales encountered in thermal systems (i.e., seconds, minutes or even longer). This guarantees that the processes, albeit irreversible, can run in a quasistatic way and pass through a series of quasi-equilibrium states with well-defined thermodynamic potentials and state variables of the bodies. This is in marked contrast to non-equilibrium oscillatory processes such as the Belousov-Zhabotinski chemical reaction (*21,22*), other oscillations in complex systems far from thermodynamic equilibrium (*23*), to "thermal-inductor" type of behaviors associated with the convection of heated fluids (*24*), or to transient switching operations in light-emitting diodes (*25*).

**Experiments**

In a real experiment, measurements of sizeable temperature oscillations face certain challenges. At present, the most efficient Peltier elements have a maximum $Z_T \approx 2$ (*26*). In a scenario of cooling an amount of boiling water from 100 ºC down to its freezing temperature at 0 ºC by connecting it to a heat sink at room temperature (20 ºC), for example, a $Z_T \approx 5$ would be required (Fig. 3A), which is out of reach of the present technology. A further challenge is the choice of the inductance $L$. It should be large enough to allow for the cooling of a substantial amount of material while keeping $k$ as small as possible to maximize $Z_T$. With $\tau^* = C/k$ of the order of several seconds and typical internal resistivities of Peltier elements $R \approx 0.1$ Ω and higher, $L_{opt} > L^* = R\tau^*$ must be of the order of 1 H or larger, although a useful cooling effect could still be achieved for $L < L^*$. Normal conducting inductors with such large inductances suffer from a considerable internal resistance $R_s$, however, thereby reducing the cooling performance well below the theoretical expectations. The incorporation of a corresponding finite resistivity $R_s$ switched in series, in addition to the internal resistance $R$ of the Peltier element as it is sketched in the inset of Fig. 4A, leads again to Eq. 3 but with $R$ replaced by $R + R_s$ (see Supplementary Materials, section S3). This will reduce the effective dimensionless figure of merit $Z_T$ of the Peltier element by a factor $R/(R + R_s)$, which can be substantial as soon as $R_s$ becomes of the order of $R$ or even larger, with an associated decrease of the cooling performance as it is illustrated in Fig. 3A and in section S1 of the Supplementary Materials.

Therefore, we performed an experiment with two configurations of resistanceless superconducting coils (with $L$ = 30 H and 58.5 H, respectively), connected to a commercially available Peltier element. With a cube of ≈ 1 cm³ copper as the thermal load at temperature $T_b$ and a massive copper base acting as the thermal reservoir held at $T_r$ = 295 K (≈ 22 °C), we are thereby implementing an *entirely passive* oscillating thermal circuit that should show the predicted temperature oscillations (see Materials and Methods). In Fig. 4A we present the results of these measurements according to the experimental scheme shown in Fig. 1B and analyzed in Fig. 3A. Initially heated to 377 K (≈ 104 ºC), $T_b$ dropped for $L$ = 58.5 H by ≈ 1.7 K below the base temperature $T_r$, clearly verifying that heat has indeed been flowing from the chilling copper cube to the thermal reservoir as soon as $T_b$ fell below 295 K around $t_0 \approx 410$ s. Using the known values $k$, $R$ and $\alpha$ of the Peltier element that we had previously obtained by the procedure described further below, we modeled these experimental $T_b(t)$ data by taking into account the parasitic resistance $R_s$ in series due to the long electric lines to the superconducting coils (inset of Fig. 4A), according to the corresponding analogon to Eq. 2 (see Eq. S5 in Section S3 of the Supplementary Materials) and Eq. 1a . The data are best reproduced with $R_s$ = 21 mΩ and 43 mΩ for $L$ = 30.0 H and 58.5 H, respectively (solid lines in Fig. 4A), in very good agreement with our direct measurements for the total resistivity of the leads and connections, $R_s \approx$ 20 mΩ and 45 mΩ (see Materials and Methods).

In Fig. 5 we have also plotted the different fractions of the rates of energy flow in this experiment, as we have depicted them in Fig. 2B. The corresponding data were derived from the measured temperature difference $\Delta T = T_b(t) - T_r$, the electric current $I$, and the known parameters of the Peltier element. Except for the comparably small electromagnetic power term ($LI\dot{I}$) and the parasitic Joule heating along $R_s$, all the relevant exchange of energy occurs directly between the Peltier element, the thermal load and the thermal bath, with the thermoelectric contributions $\alpha T_b I$ and $\alpha T_r I$ clearly dominating the other terms appearing in Eqs. 1b and 1c.

In order to be better able to quantitatively analyze our model and to precisely determine the parameters of the Peltier element, we had previously performed a series of additional experiments using an active gyrator-type substitute of a real inductor (*27*) that allows simulating almost arbitrary values of $L$ with negligible effective internal resistivity, $R_s \approx 0$. Although the thermal connection can then no longer be considered as strictly operating "without external intervention", no net external work is performed on the system. In Fig. 4B we present the results of a series of corresponding measurements, for four different values of the nominal inductance $L$ of the gyrator ($L$ = 33.2 H, 53.6 H, 90.9 H and 150 H), with $T_b$ reaching a temperature ≈ 2.7 K below $T_r$ for $L$ = 90.9 H. In the same figure, we also show the results of a global fit to both the measured $T_b(t)$ and $I$(t) data according to the relations given in this work, with only four free parameters, $C, R, k$ and $Z_T$. We obtain an excellent agreement for an effective $Z_T$ = 0.432, $C$ = 4.96 J/K (this value includes a heat-capacity contribution of the Peltier element), $R$ = 0.22 Ω, and $k$ = 0.0318 W/K, resulting in a Seebeck coefficient $\alpha = \sqrt{Z_T k R / T_r}$ = 3.21x10$^{-3}$ V/K. The used values $L$ = 33.2 H and 90.9 H are very close to the resulting $L^*$ = 34.3 H and the optimum inductance $L_{opt}$ = 94.4 H, respectively (see Supplementary Materials, section S1).

**Relation to the second law of thermodynamics**
The fact that heat temporarily flows from cold to hot in the processes described above inevitably calls for a further discussion in view of the second law of thermodynamics. The proof that these processes do not violate this fundamental law is surprisingly simple. The total rate of entropy production is $\dot{S}_{tot} = \dot{S}_b + \dot{S}_r = \dot{Q}_b / T_b + \dot{Q}_r / T_r$. The empty current-carrying inductor, which can be placed remote from the Peltier element to inhibit any exchange of heat, does not contribute at all to the entropy balance, and the associated magnetic contribution to the entropy also vanishes because the corresponding Gibb's free energy does not depend on the temperature. While the terms in the Eqs. 1b and 1c related to the Peltier element cancel out, the corresponding contributions of the internal resistivity are $\frac{1}{2}RI^2(1/T_b + 1/T_r) > 0$, those due to heat conduction $-k(T_b - T_r)/T_b + k(T_b - T_r)/T_r \geq 0$, and we have indeed $\dot{S}_{tot} > 0$. The temperatures $T_b(t)$ and $T_r(t)$ counter-oscillate in the second scenario (Figs. 1D and 3B) and repeatedly match around $\bar{T}$, which seems to be at odds with the expectation that the total entropy of the two bodies with $T_b = T_r$ must be larger than with $T_b \neq T_r$. As $T_{av}$ is not constant but lowest for $T_b \approx T_r < \bar{T}$ where the small but finite amount of magnetic energy is largest, the total change in entropy $\Delta S_{tot}(t)$ is indeed a monotonously increasing function of time $t$ (inset of Fig. 3B). For $t \to \infty$ and in the limit $\Delta_0 \ll \bar{T}$, it amounts to $\Delta S_{tot}/2C \approx (\Delta_0/2\bar{T})^2$, the same value as we would have in a corresponding experiment without a Peltier-*LCR* circuit.

Although no external intervention is generating the inversion of the temperature gradients and forcing heat to flow from cold to hot, the inductor is, as an integrated part of the thermal connection, perpetually changing its state due to the oscillatory electric current. Therefore, the described processes are also in full conformity with the original postulate of the second law of thermodynamics by Clausius, stating that a flow of heat from cold to hot must be associated with "some other change, connected therewith, occurring at the same time" (*16*).

We may finally analyze the considered thermal oscillator in view of the thermodynamic efficiency of heat engines. If we assume an ideal Carnot efficiency (*15*) for the stage (*i*) in the configuration of Fig. 1B where the circuit is acting as a thermoelectric generator, a maximum amount of work $dW \leq CdT_b(1 - T_r/T_b)$ can be extracted from the heat capacity $C$ by successively removing infinitesimally small amounts of heat $CdT_b$ upon cooling it by $dT_b$, summing up to $W \leq C(\Delta_0 - T_r \ln[T_b(0)/T_r])$ for the full range of temperatures from $T_b(0) = T_r + \Delta_0$ down to $T_r$. By completely recycling this amount of work during stage (*ii*), where the thermal load is chilled from $T_b = T_r$ further down with a corresponding Carnot engine acting as a cooler, we reach the lowest possible temperature $T_{b,\min}$ that can be achieved in this way, which is given by the implicit equation

$$T_r \ln\left(\frac{T_b(0)}{T_{b,\min}}\right) = T_b(0) - T_{b,\min} \tag{5}$$

(see section S1 of the Supplementary Materials and Fig. S1C therein). This result coincides with Eq. 4 in the limits $Z_T \to \infty$ and $\Delta_0 \ll T_r$, where $T_r - T_{b,\min} \to \Delta_0$. It is independent of any model assumptions, and therefore represents an ultimate limit for the maximum possible cooling effect during any thermal oscillation cycle without external work done on a system. However, this thermodynamic constraint is neither crucial for the experiments reported here, nor for practical applications using present technology with $Z_T$ values still of the order of unity (*26*). In the temperature range used for the present experiments, for example, the lowest achievable temperature allowed by thermodynamics could still be as low as $T_{b,\text{min}} \approx 226$ K ($\approx -47$ °C).

**CONCLUSIONS**

We have shown both theoretically and experimentally that the use of a thermal connection containing a variant of a "thermal inductor" can drive the flow of heat from a cold to a hot object without external intervention. We have analyzed a corresponding oscillatory thermal process in detail, both from the electrical and the thermodynamic point of view. While it fully complies with the second law of thermodynamics, we identify a general thermodynamic limit for the maximum possible cooling effect that can occur during such a thermal oscillation cycle. Despite the conceptual simplicity of the described experiment and based on the laws of classical physics, it has, to our knowledge, never been considered in the literature. With future progress in materials research, the technique may become technically useful and allow for the cooling of hot materials well below the ambient temperature without the need for an external energy supply or any moving parts.

**MATERIALS AND METHODS**

We used a commercial Peltier element (Kryotherm Inc., Module TB-7-1.4-2.5) for all the experiments. The temperature difference between the thermal load (a copper cube of dimensions $\approx 10\text{x}10\text{x}10$ mm$^3$ with a mass m $\approx$ 9 g) and the thermal bath (a block of 3.65 kg copper) was measured with a copper-constantan differential thermocouple, thermally connected to the experiment with silver paint. The corresponding voltage was measured using a Keysight multimeter 34465 A, and converted to a temperature difference according to a standard type-T conversion table.

    The superconducting coils used in the experiment are part of two PPMS experimental platforms (Physical Property Measurements System, Quantum Design Inc.), immersed in liquid helium at $T = 4.2$ K. The resistivities $R_s$ of the electric lines and connections to these coils were measured with a Peaktech 2705 milliohm meter with an accuracy of +/- 2 mΩ. For one superconducting coil with inductance $L = 30$ H we measured $R_s = 20$ mΩ, and for two coils in series with a total $L = 58.5$ H we obtained $R_s = 45$ mΩ. The inductances were determined with a $\approx$ 2 % accuracy by measuring the time constant of a decaying current in a closed loop with a known 20 Ω resistor. The resistivities $R_{hs} \approx 35$ Ω of the heat-switches ("persistent switches") in parallel to the superconducting coils have been included in this analysis. For the operation in a thermal oscillator experiment, their presence is quantitatively negligible, however.

    The gyrator was built based on the scheme described in (*27*). The values for *L* can be adjusted by an appropriate choice of a resistor in the gyrator circuit, and they have been crosschecked by measuring the time constants of a respective *LR* test circuit. The current *I* was

determined by monitoring the voltage across an internal shunt resistor inside the gyrator. All experiments were done in open air without any thermal shielding. In a global fit to all the available $T_b(t)$ and $I(t)$ data to Eq. 2 we obtain an excellent agreement for $C = 4.96(1)$ J/K (this value includes a heat-capacity contribution of the Peltier element), $R = 0.220(1)$ Ω, and effective values for $k = 0.0318(1)$ W/K and $Z_T = 0.432(1)$, resulting in $L^* = 34.3$ H and $L_{opt} = 94.4$ H. We independently confirmed the ratio of the fitted values $C$ and $k$, $\tau = C/k \approx 156.1(1)$ s, in a direct measurement by monitoring the decay of the temperature of the copper cube without the inductor connected and the Peltier circuit open, with $\tau \approx 162$ s. The effective $Z_T = 0.432$ obtained from the global fitting procedure has to be interpreted as a constant $Z_T = \alpha^2 T_r / kR$, where $k$ and $R$ in the equations (1) may include extrinsic contributions due to parasitic heat transport and internal resistance from the wiring. The true, intrinsic $Z_T$ value of the Peltier element may therefore be somewhat larger, and we estimate it to $Z_T \approx 0.5$ near room temperature. The fitted value $R = 0.22$ Ω of the circuit including the internal wiring, is also compatible with the resistance specifications of the Peltier element ($R = 0.18$ Ω +/– 10%), which represents the main source of electrical resistivity. Pictures of the experimental set-ups are provided in section S5 of the Supplementary Materials.

**SUPPLEMENTARY MATERIALS**

Section S1: Maximum possible cooling effect
Section S2: Two finite bodies with different temperature
Section S3: Effect of a finite resistance $R_s$ in series
Section S4: Analogy to a "thermal inductor"
Section S5: Pictures of the experimental set-ups
Fig. S1. Optimized values for a maximum temperature undershoot
Fig. S2. Experimental set-up of the oscillating thermal circuit

**REFERENCES**

1. N. D. Roberts and D. G. Walker, A review of thermal rectification observations and models in solid materials. *Int. J. Therm. Sci.* **50**, 648-662, (2011).
2. M. Terraneo, M. Peyrard, and G. Casati, Controlling the energy flow in nonlinear lattices: a model for a thermal rectifier. *Phys. Rev. Lett.* **88**, 094302 (2002)
3. B. Li, L. Wang, and G. Casati, Thermal diode: rectification of heat flux. *Phys. Rev. Lett.* **93**, 184301 (2004).
4. C. W. Chang, D. Okawa, A. Majumdar, and A. Zettl, Solid-state thermal rectifier. *Science* **314**, 1121–1124 (2006).
5. M. Maldovan, Sound and heat revolutions in phononics. *Nature* **502**, 209-217 (2013).
6. L. Wang and B. Li, Phononics gets hot. *Physics World* **21**, 27-29 (2008).

7. K. I. Garcia-Garcia and J. Alvarez-Quintana, Thermal rectification assisted by lattice transitions. *Int. J. Therm. Sci.* **81,** 76-83 (2014).
8. P. Ben-Abdallah, S. A. Biehs, Phase-change radiative diode. *Appl. Phys. Lett.* **103**, 191907 (2013).
9. E. Pallecchi, Z. Chen, G. E. Fernandes, Y. Wan, J. H. Kim and J. Xu, A thermal diode and novel implementation in a phase-change material. *Mater. Horiz.* **2**, 125-129 (2015).
10. A. L. Cottrill and M. S. Strano, Analysis of Thermal Diodes Enabled by Junctions of Phase Change Materials. *Adv. Energy Mater.* **5**, 1500921 (2015).
11. R. Scheibner, M. König, D. Reuter, A. D. Wieck, C. Gould, H. Buhmann, and L. W. Molenkamp, Quantum dot as thermal rectifier. *New J. Phys.* **10**, 083016 (2008).
12. M. J. Martínez-Pérez, A. Fornieri, and F. Giazotto, Rectification of electronic heat by a hybrid thermal diode. *Nat. Nanotechnol.* **10**, 303-307 (2015).
13. B. Li, Negative differential thermal resistance and thermal transistor. *Appl. Phys. Lett.* **88** 143501 (2006).
14. L. Wang and B. Li, Thermal logic gates: computation with phonons. *Phys. Rev. Lett.* **99**, 177208 (2007).
15. S. Carnot, Réflexions sur la puissance motrice du feu, et sur les machines propres à développer cette puissance, (Bachelier, Paris) (1824).
16. R. Clausius, Ueber eine veränderte Form des zweiten Hauptsatzes der mechanischen Wärmetheorie. *Annalen der Physik* (Poggendoff, Leipzig), **Xciii** 481-506 (1854).
17. O. Bossen and A. Schilling, LC-circuit calorimetry. *Rev. Sci. Instrum.* **82**, 094901 (2011).
18. E. Altenkirch, Elektrotechnische Kälteerzeugung. *Phys. Z.* **XII**, 920-924 (1911).
19. C. Goupil, W. Seifert, K. Zabrocki, E. Müller, and G. J. Snyder, Thermodynamics of Thermoelectric Phenomena and Applications. *Entropy* **13**, 1481-1517 (2011).
20. A. F. Ioffe, in *Semiconductor Thermoelements and Thermoelectric Cooling* (Infosearch, London, 1957).
21. B. P. Belousov, in *Oscillations and Travelling Waves in Chemical Systems*, R. J. Field and M. Burger. Eds. (Wiley, New York, 1985), p. 605-613.
22. A. M. Zhabotinsky, Periodical oxidation of malonic acid in solution (a study of the Belousov reaction kinetics). *Biofizika* **9**, 306-311 (1964).
23. G. Nicolis and I. Prigogine, in *Self-Organization in Non-equilibrium Systems: From Dissipative Structures to Order Through Fluctuations* (Wiley, New York, 1977).
24. R. C. L. Bosworth, Thermal inductance. *Nature (London)* **158**, 309 (1946).
25. H. Ye, S. Y. Y. Leung, C. K. Y. Wong, K. Lin, X. Chen, J. Fan, S. Kjelstrup, X. Fan, and G. Q. Zhang, Thermal Inductance in GaN Devices. *IEEE Electron. Device Lett.* **37**, 1473-1476 (2016).
26. T. Tritt, and M. Subramanian, Thermoelectric Materials, Phenomena, and Applications: A Bird's Eye View. *MRS Bull.* **31**, 188-194 (2006).
27. R. H. S. Riordan, Simulated inductors using differential amplifiers. *Electron. Lett.* **3**, 50-51 (1967).

**Acknowledgments:** We thank to Dr. Achim Vollhardt, Daniel Florin and David Wolf for their technical support. **Funding:** The authors acknowledge the financial support by the UZH and by the Schweizerische Nationalfonds zur Förderung der Wissenschaftlichen Forschung (Grant. No. 20-131899). **Author contributions:** O.B. introduced the concept of using a Peltier element in combination with an inductor for thermal experiments; X.Z modified and characterized the set-up for the use with the superconducting coils; A.S. wrote the manuscript, performed all the measurements and the full analysis of the circuit with regard to its use as passive thermal oscillator and its compliance with thermodynamics. **Competing interests:** The authors declare that they have no competing interests. **Data and materials availability:** All data needed to evaluate the conclusions in the paper are present in the paper and/or the Supplementary Materials. Additional data related to this paper may be requested from the authors.

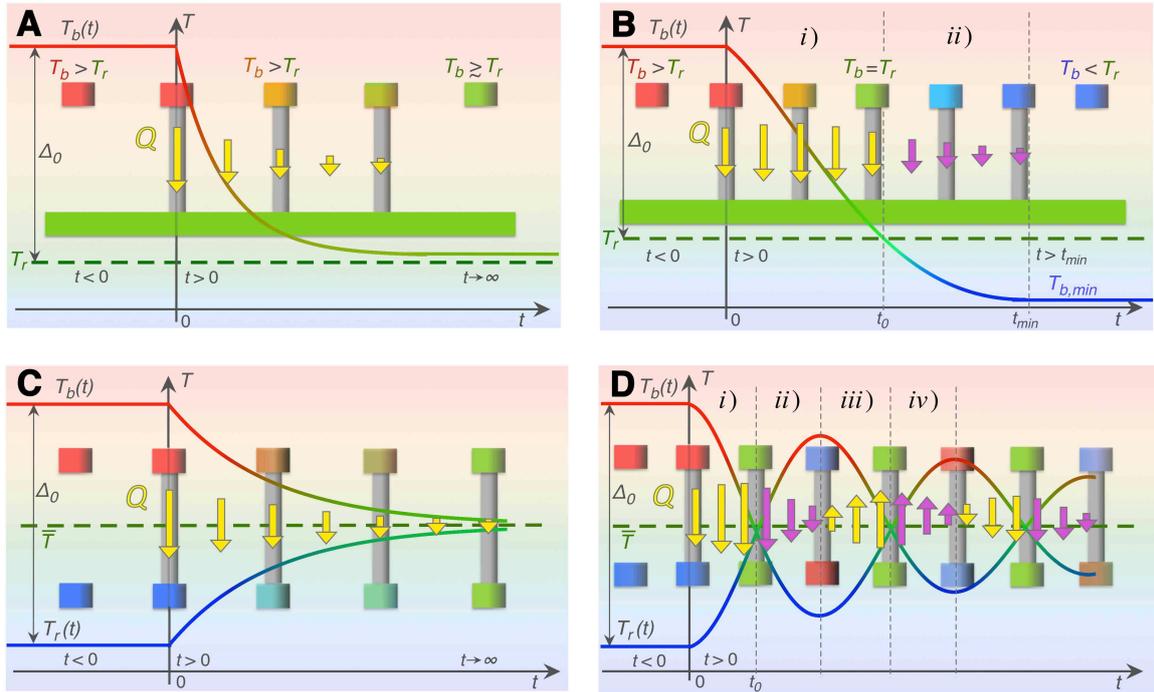

**Fig. 1. Sketch of two situations in which objects with different temperatures are thermally connected.** Arrows represent the direction of the flow of heat from (light/yellow) or to (dark/purple) the respective warmer object. (**A**) When an initially hot body is thermally connected at time $t = 0$ to a colder thermal reservoir held at temperature $T_r$, its temperature $T_b$ is expected to drop monotonically by the loss of heat $Q$ to the colder reservoir, and to approach $T_r$ in the limit $t \to \infty$. (**B**) Sketch of a process in which $T_b$ undershoots the temperature of the reservoir for $t > t_0$, and heat $Q$ is thereafter temporarily transferred from the chilling body to the warmer reservoir. The lowest temperature of the body $T_{b,\min} < T_r$ is reached at $t = t_{\min}$ when the connection can be removed. (**C**) Two similarly connected finite heat capacities are expected to smoothly approach thermodynamic equilibrium at a mean temperature $\bar{T}$, with heat flowing in one direction only and always $T_b > T_r$. (**D**) Two bodies showing opposite oscillations in temperature, with an alternating direction of the heat flow and a repeated temporary transfer of heat from cold to hot. The roman numerals (*i*) – (*iv*) refer to the four quarters of the period of one full oscillation cycle of $T_b(t)$ as elaborated in the text and in the caption of Fig. 2B.

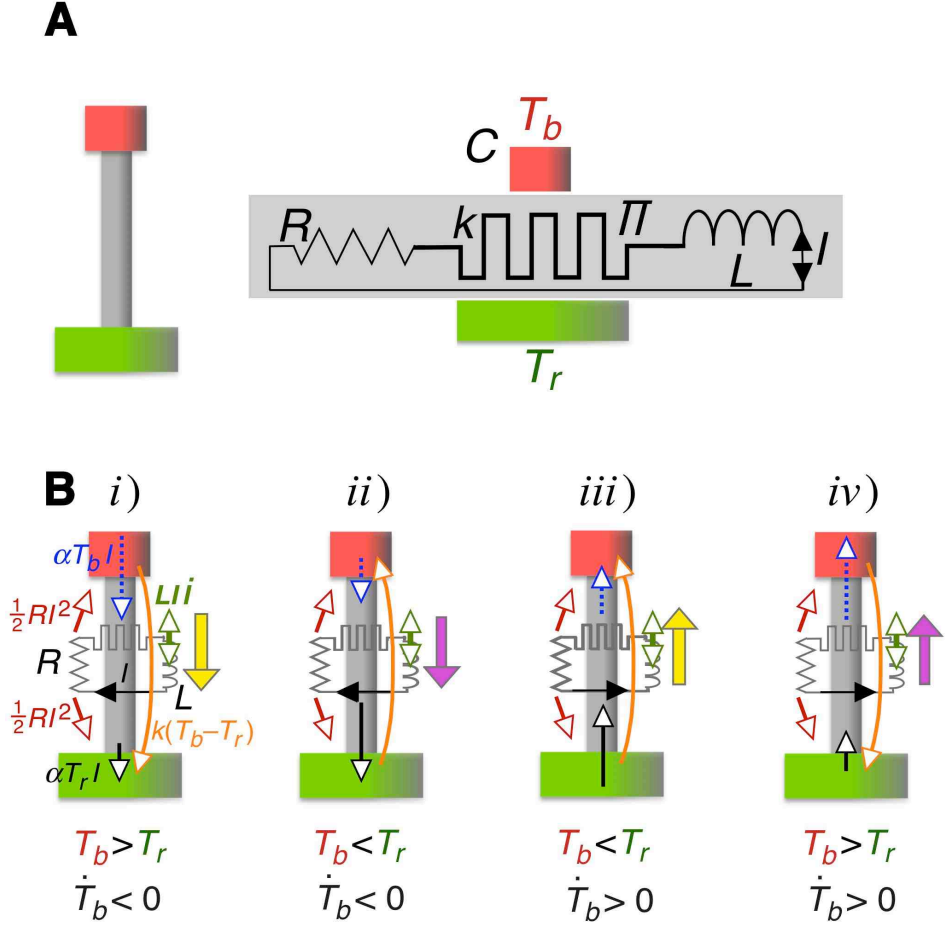

**Fig. 2. Equivalent electrical network and illustration of the heat flow within the considered thermal connection between a body with heat capacity $C$ at temperature $T_b$ and another body or a thermal reservoir at $T_r$.**

**(A)** The electrical network consists of a Peltier element ($\Pi$) with internal resistance $R$ and thermal conductance $k$, in a closed circuit with an ideal inductance $L$. The oscillatory current $I$ is ultimately driven by the voltages supplied by the thermoelectric effect due to the temperature difference between the cold and the hot end of the Peltier element, and the induced voltage $L\dot{I}$ across $L$, see Eq. 1a.

**(B)** Sketch of the individual contributions to the flow of heat (open arrowheads, arrow lengths not to scale) in Eqs. 1b and 1c for situations when heat is flowing from (filled light/yellow arrows) or to (filled dark/purple arrows) the warmer end of the Peltier element, drawn for one oscillation cycle of $T_b(t)$ as depicted in Figs. 1B and 1D. The thermal oscillator acts during a full period of an oscillation cycle of $T_b(t)$ alternately as a thermoelectric generator (*i*), as a cooler (*ii*), a generator (*iii*), and as a thermoelectric heater (*iv*). During all these processes, a small amount of electromagnetic power ($L I \dot{i}$, green double arrows) is exchanged with the inductor although the total stored magnetic energy $\tfrac{1}{2} L I^2$ is always less than a fraction $\Delta_0 / T_r$ of the initially deposited excess heat $\sim C \Delta_0$ (see text and Fig. 5).

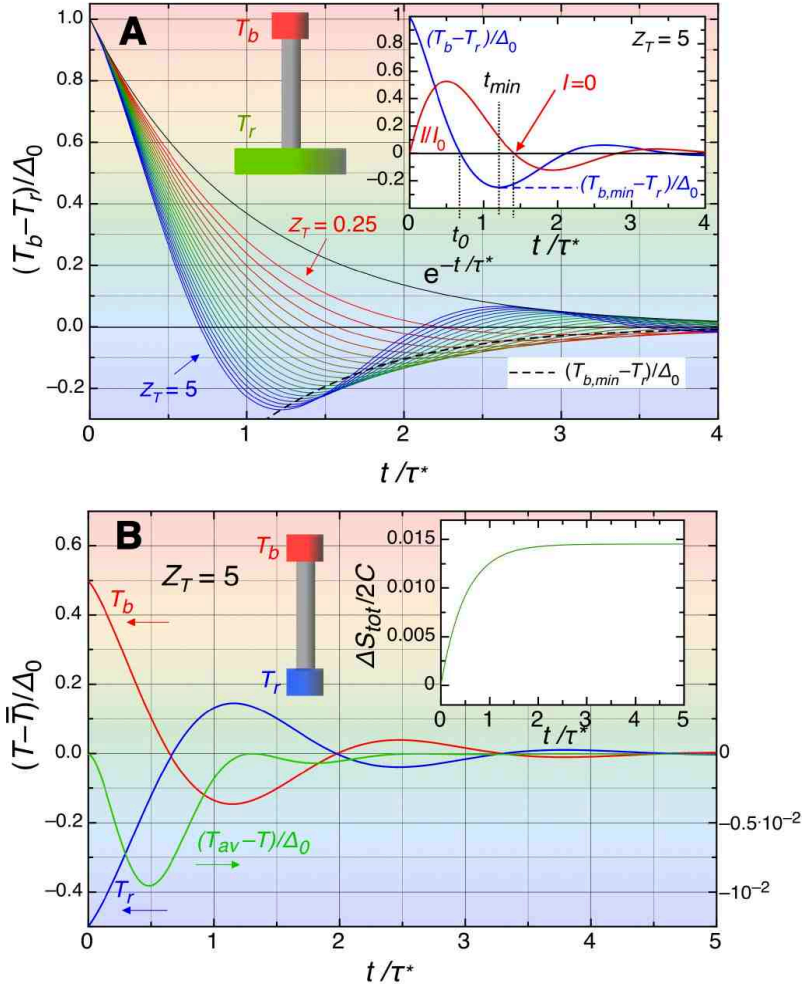

**Fig. 3. Evolution of the temperature difference between a cooling body and a thermal bath or another finite body, which are connected in an experiment using a "thermal inductor".**

(**A**) Normalized temperature difference $(T_b(t) - T_r)/\Delta_0$ between a finite body and a thermal reservoir for $L = L^* = RC/k$ and $Z_T$ between 0.25 (red) and 5 (blue) in steps of 0.25, obtained from solving Eq. 3. The time is in units of $\tau^* = C/k$. The black line represents a corresponding relaxation process with a time constant $\tau^*$, which would take place if the Peltier circuit were interrupted from the beginning. If the thermal connection is not removed after reaching the respective $T_{b,\min}$ (dashed line), $T_b(t)$ approaches thermal equilibrium with eventually $T_b = T_r$ in all cases. The inset shows the damped oscillations of both $T_b(t)$ and $I(t)$.

(**B**) Temperatures $T_b(t)$ and $T_r(t)$ of two connected finite bodies with equal heat capacities, relative to the mean initial temperature $\bar{T} = [T_b(0) + T_r(0)]/2$ and normalized to the initial temperature difference $\Delta_0$, for $Z_T = 5$ (time in units of $\tau^*$). $T_{av}$ denotes their average value showing local minima around $T_b \approx T_r$ (the numbers for $T_{av}$ were calculated for $\Delta_0/T_r = 0.27$). The inset shows the evolution of the total entropy gain as a function of time in corresponding normalized units.

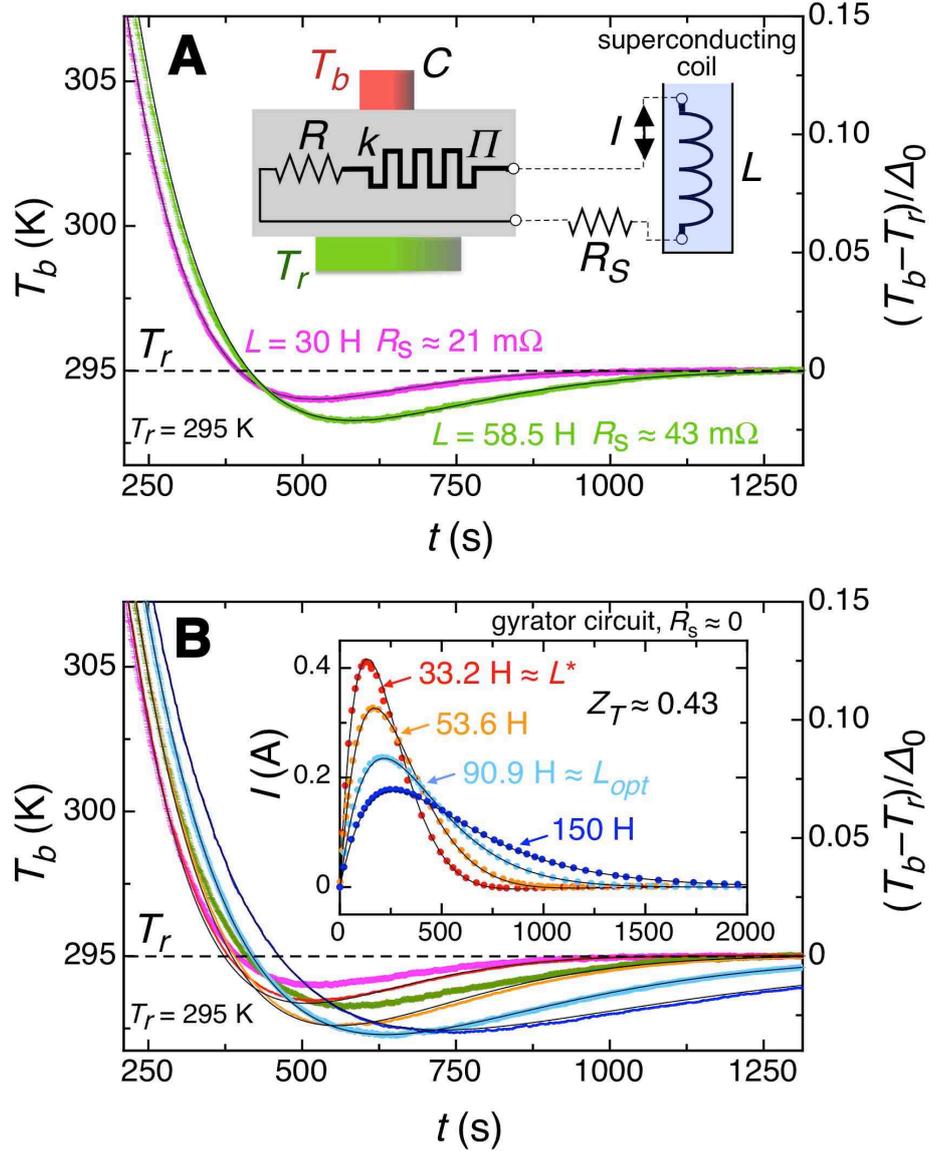

**Fig. 4. Results from experiments with oscillating thermal circuits containing the equivalent of a "thermal inductor".**

(**A**) Temperature $T_b(t)$ data taken for two configurations of superconducting coils with $L = 30$ H and 58.5 H, respectively. In this type of an experiment, the oscillating thermal circuit is entirely passive. The temperature $T_b(t)$ of a copper cube that has been thermally connected to a heat reservoir held at $T_r = 295$ K and initially heated by $\Delta_0 = T_b(0) - T_r \approx 82$ K significantly undershoots with respect to $T_r$ by $\approx 1.7$ K for $L = 58.5$ H. The inset shows the respective equivalent electrical network, including a parasitic electrical resistance $R_s$ in series due to electrical leads and connections. The solid lines are $T_b(t)$ data obtained by solving the corresponding relevant differential equations using the parameters from a global fit to the data shown in Fig. 4B, and with $R_s = 21$ mΩ and 43 mΩ for $L = 30.0$ H and 58.5 H, respectively.

(**B**) Experiments using a gyrator-type substitute of an electric inductor with a nominal $R_s \approx 0$. Main panel: Temperature $T_b(t)$ for four different values of nominal inductance $L$, with a maximum undershoot of $T_b(t)$ with respect to $T_r$ by $\approx 2.7$ K for $L = 90.9$ H. The $T_b(t)$ data from Fig. 4A (green and purple dots) are included for comparison. Inset: Evolution of the electric current flowing through the Peltier element. The solid black lines correspond to a global fit to the four data sets according to the relations given in the main text, with the fitting parameters $C$, $R$, $k$ and $Z_T$.

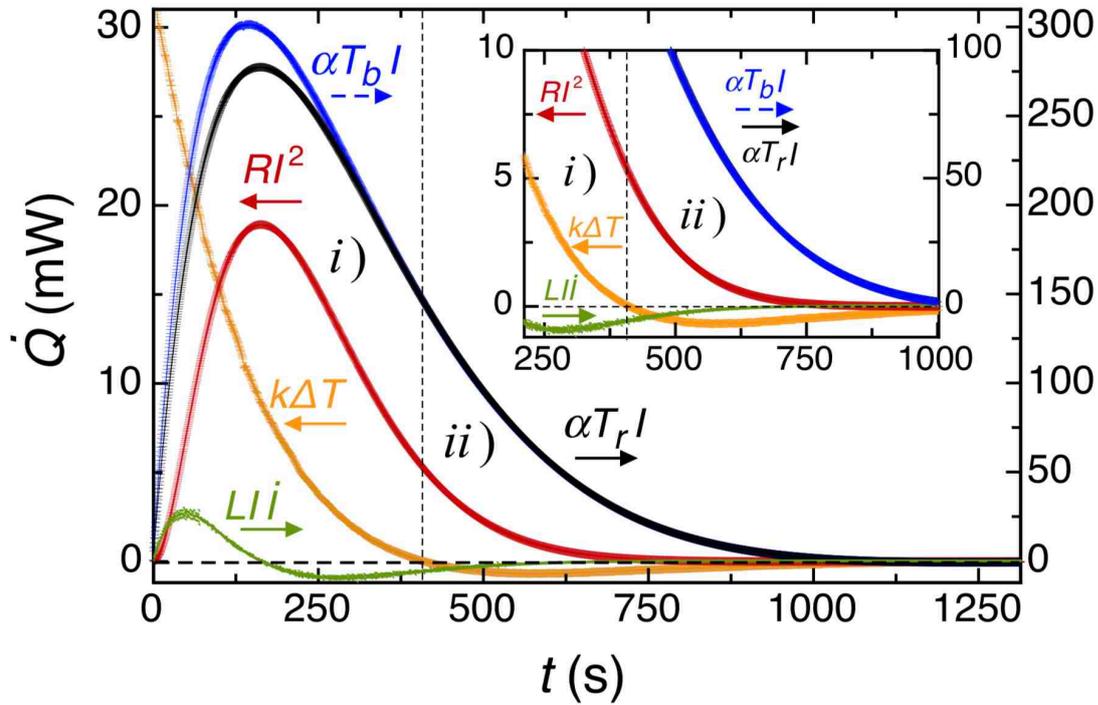

**Fig. 5. The different fractions of the rates of energy flow in the oscillating thermal circuit for the experiment with $L$ = 58.5 H shown in Fig. 4A.** The roman numerals (*i*) and (*ii*) refer to the definition provided in the text and in the caption of Fig. 2B. The subsequent stages (*iii*) and (*iv*) are not discernible in these data due to the still quite low value $Z_T$ = 0.432 (see also Fig. 3A). The thermoelectric contributions $\alpha T_b I$ and $\alpha T_r I$ (right scales) clearly dominate all the other terms in Eqs. 1b and 1c (left scales).

## Supplementary Materials

### Section S1: Maximum possible cooling effect

The maximum possible temperature difference $|T_b - T_r|$ for $t > 0$ and for a given value of $Z_T$ is not fully reached for the parameters used in the main text, but with $L_{opt} = \lambda L^*$ where $\lambda(Z_T) > 1$. The numerically obtained function $\lambda(Z_T)$ is shown in Fig. S1A for $\Delta_0 = T_b(0) - T_r \ll T_r$. In the limit $Z_T \to \infty$, we have $L_{opt} = L^*$. For a finite $Z_T$, however, taking $L = L_{opt}$ results in $\omega_{opt} = \omega^* \sqrt{(Z_T + 1)/(\lambda Z_T) - (\lambda + 1)^2/(4\lambda^2 Z_T)}$ and $\tau_{opt} = \tau^* 2\lambda/(1+\lambda)$, and a truly oscillatory behaviour of $I(t)$ is possible only for $Z_T > 1/3$. Even if the solutions for $I(t)$ are over-damped for $Z_T < 1/3$, the temperature of the body still undershoots with the initial conditions used in the main text for all values of $Z_T$. A choice of $L = L_{opt}$ instead of $L^*$ leads only to a moderately improved cooling performance of the device, however, at the cost of significantly increasing the required inductance $L$ (see Fig. S1B).

A general thermodynamic limit for the maximum cooling effect of any oscillating thermal circuit without involving external work can be obtained by assuming ideal Carnot efficiency for the whole process. During stage (*i*) in the set-up shown in Fig. 1B, a maximum amount of work $dW \leq CdT_b(1 - T_r/T_b)$ can be extracted by removing an infinitesimally small quantity of heat $CdT_b$ upon cooling the heat capacity $C$ by $dT_b$. Integrating over the range of temperatures $T_b$ from $T_b(0) = T_r + \Delta_0$ down to $T_r$ yields

$$W \leq C\left(\Delta_0 - T_r \ln\left(\frac{T_b(0)}{T_r}\right)\right). \tag{S1}$$

When further cooling it down from $T_r$ to $T_{b,\min}$, the device must also obey the Carnot limit for a cooling engine, where the maximum amount of removed heat $CdT_b$ with respect to the external work $dW'$ is $CdT_b \leq dW' T_b/(T_r - T_b)$. By integration from $T_{b,\min}$ to $T_r$ we obtain

$$W' \geq C\left(T_r \ln\left(\frac{T_r}{T_{b,\min}}\right) - T_r + T_{b,\min}\right). \tag{S2}$$

Combining the two processes by recycling the work $W$ harvested during stage (*i*) to 100% and using it for the subsequent cooling of $C$ in stage (*ii*) yields with $W = W'$ an implicit equation for $T_{b,\min}$ (see Fig. S1C),

$$T_r \ln\left(\frac{T_b(0)}{T_{b,\min}}\right) = T_b(0) - T_{b,\min}. \tag{S3}$$

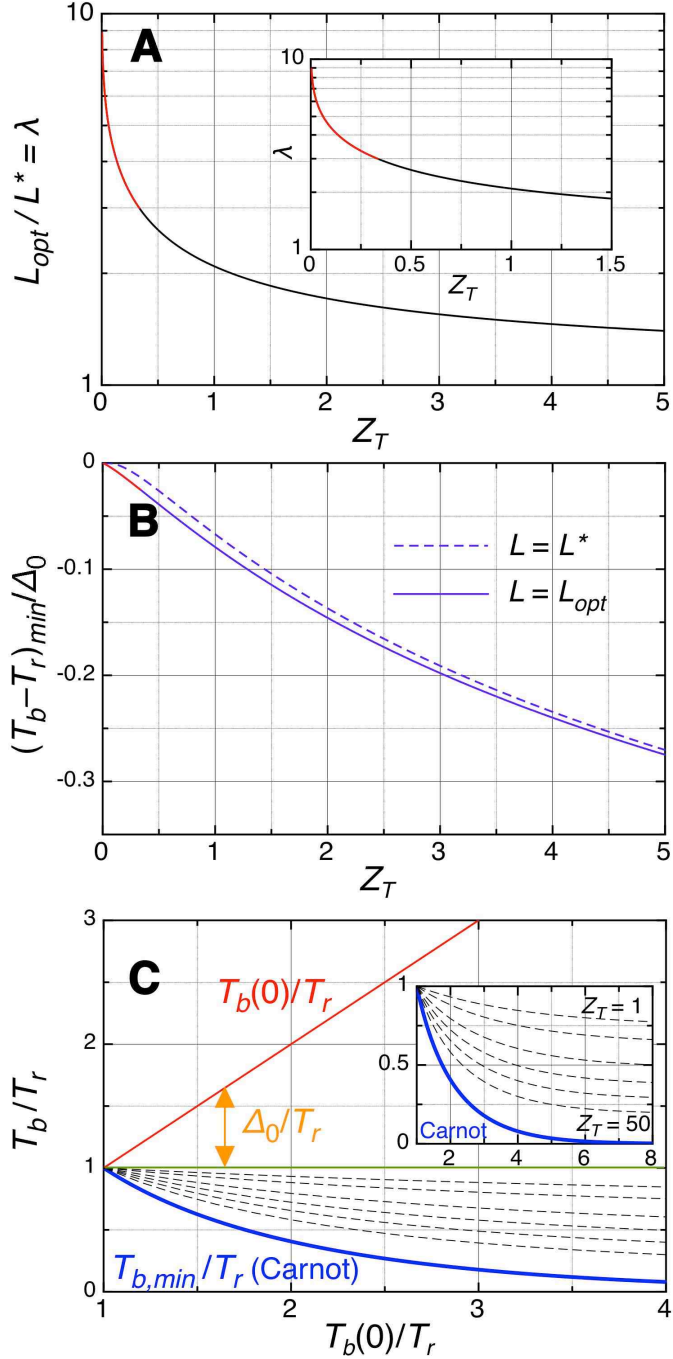

**Fig. S1. Optimized values for a maximum temperature undershoot**. (**A**) Optimum inductance $L_{opt} = \lambda(Z_T)RC/k$, in units of $L^* = RC/k$. If $L = L_{opt}$, $T_b$ exhibits the maximum possible undershoot with respect to the temperature $T_r$. (**B**) Temperature difference $(T_b - T_r)/\Delta_0$ relative to the initial value $T_b(0) - T_r = \Delta_0 \ll T_r$, for $L = L_{opt}$ (solid line) and $L = L^*$ (dashed line). While in the latter case, $I(t)$ shows a true oscillation for all values of $Z_T$, the current $I(t)$ is over-damped for $Z_T < 1/3$ when choosing $L = L_{opt}$ but still results in an undershoot of $T_b(t)$ (red lines). (**C**) Lowest possible reduced temperature $T_{b,\min}/T_r$ allowed by thermodynamics (Carnot limit, solid blue line), as a function of the reduced starting temperature $T_b(0)/T_r$. The dashed lines are corresponding $T_{b,\min}/T_r$ data for $Z_T$ = 1, 2, 5, 10, 20, and 50 (top to bottom), obtained by numerically solving Eq. 2 with the respective $L = L_{opt}$.

### Section S2: Two finite bodies with different temperature

If we ascribe a heat capacity $2C$ to each of the bodies, the corresponding amounts of exchanged heat are $\dot{Q}_b = 2C\dot{T}_b$ and $\dot{Q}_r = 2C\dot{T}_r$, respectively. Combining these quantities with the Eqs. 1 by eliminating both $T_b$ and $T_r$, we obtain the nonlinear differential equation for $I(t)$,

$$LC\ddot{I} + (RC + kL)\dot{I} + (kR + \alpha^2\bar{T})I - \alpha^2 LI^3/8C = 0, \qquad (S4)$$

with the mean initial temperature $\bar{T} = [T_b(0) + T_r(0)]/2$. The time dependent temperatures $T_b(t)$ and $T_r(t)$ can again be obtained from the Kirchhoff relation (1a). Additional conditions are the thermal balance equation and the energy conservation law, requiring $2\bar{T} = T_b + T_r + LI^2/4C$ (which can also be derived from the Kirchhoff equation), and $2\bar{T} = T_b + T_r$ for $t \to \infty$ and $I \to 0$ if the two heat capacities are assumed to be temperature independent.

In the limit $\Delta_0 = T_b(0) - T_r(0) \ll \bar{T}$, the amount of magnetic energy stored in the inductor is vanishingly small, as we have shown in the main text, i.e., $\frac{1}{2}LI^2 \ll 4C\bar{T}$. Therefore we may neglect the last term in Eq. S4 and end up in this limit with the same differential equation Eq. 3 for $I(t)$ of a damped harmonic oscillator but with $T_r$ replaced by $\bar{T}$.

### Section S3: Effect of a finite resistance $R_s$ in series

In a real experiment, cables and/or the use of a normal conducting coil will contribute to the circuit with a finite electrical resistivity $R_s$ in series in addition to the internal resistance $R$ of the Peltier element (see inset of Fig. 4A). In the case of a finite heat capacity connected to a thermal reservoir, Eq. 2 then becomes

$$LC\ddot{I} + ((R + R_s)C + kL)\dot{I} + (k(R + R_s) + \alpha^2 T_r)I + \alpha(\tfrac{1}{2}R + R_s)I^2 + \alpha LI\dot{I} = 0. \quad (S5)$$

If we again restrict ourselves to $\Delta_0 = T_b(0) - T_r \ll T_r$ the last two terms are negligible because with Eq. 1a, $(\tfrac{1}{2}R + R_s)I^2 + LI\dot{I} < (R + R_s)I^2 + LI\dot{I} = \alpha(T_b - T_r)I \ll \alpha T_r I$, resulting in the same Eq. 3 for a damped harmonic oscillator, but with the $R$ replaced by $R + R_s$.

### Section S4: Analogy to a "thermal inductor"

Thermal circuits can be mapped onto electrical circuits by replacing the thermal conductance with an electrical conductance, heat capacities with electrical capacitances, and temperature differences with voltage differences $\Delta V$. The resulting differential equations for the heat $Q$ and heat currents $I_{th} = \dot{Q}$, or charge and electric currents $I$, respectively, turn out to be equivalent.

A corresponding "thermal inductor" (with thermal self-inductance $L_{th}$) would have, in a strict sense, to fulfill the proportionality between the time derivative of the thermal current $\dot{I}_{th} = \ddot{Q}$ and the resulting temperature difference $\Delta T$ in an analogous way to $L\dot{I} = -\Delta V$, so that $L_{th}\dot{I}_{th} = -\Delta T$, where the unit of $L_{th}$ is 1 Ks$^2$/J.

We consider the case of a finite body in thermal contact with an infinite thermal reservoir as shown in Figs. 1A and 1B in the main text. We again assume that the temperature variations are sufficiently small so that $|T_b(0) - T_r| = \Delta_0 \ll T_r$, and that we have perfect electrical conductors with $R = 0$ and negligible Joule heating. In analogy to an ideal resistanceless electric coil, we also assume perfect thermal insulation between both ends of the "thermal inductor", i.e., with no thermal leakage current in parallel due to a finite thermal conductance $k$. We then obtain with Eqs. 1a, 1b, $k = 0$, $R = 0$ and $\Delta T = T_b - T_r$

$$\dot{I}_{th} = \ddot{Q}_b = -\alpha \dot{T}_b I - \alpha T_b \dot{I} = -\alpha \dot{T}_b I - \alpha^2 T_b \Delta T / L. \qquad (S6)$$

The analogy $L_{th}\dot{I}_{th} = -\Delta T$ holds if the first summand on the right side of Eq. S6 is negligible, and we then have (17)

$$L_{th} \approx L/(\alpha^2 T_b) \approx L/(\alpha^2 T_r). \qquad (S7)$$

To fulfill this condition for an arbitrary variation of the electric current with time, $I(t) = I_0 f(t)$, we must require $|\dot{T}_b / T_b| \ll |\dot{I}/I|$, or in terms of $f(t)$,

$$\frac{\Delta_0}{T_r} \left|\frac{\ddot{f}(t)}{\dot{f}(0)}\right| \ll \left|\frac{\dot{f}(t)}{f(t)}\right|. \qquad (S8)$$

For $\Delta_0 \ll T_r$ this is fulfilled all the time in most conceivable cases where the variations in $T_b(t)$ are of the order of $\Delta_0 \ll T_r$, e.g., for $f(t) = \gamma t$, $f(t) = \exp(-t/\tau)$, or $f(t) = 1 - \exp(-t/\tau)$. In this sense, it is justified to interpret the present circuit as the equivalent of a "thermal inductor". For oscillatory functions where $\dot{f}(t)$ temporarily vanishes while $\ddot{f}(t)$ remains finite as in our case of a damped oscillation of $f(t)$, the condition Eq. S8 is momentarily violated, but it is still respected most of the time as a result of the exponential decay of the oscillating electric current, and $\Delta_0 \ll T_r$.

In this sense, we can re-write Eq. 3 in the main text in terms of only thermal quantities, e.g., for the thermal current $I_{th} = \dot{Q}_r$:

$$L_{th}C\ddot{I}_{th} + (\frac{C}{Z_T k} + kL_{th})\dot{I}_{th} + (\frac{1}{Z_T} + 1)I_{th} \approx 0. \qquad (S9)$$

It is analogous to that of an electric LCR circuit with two dissipative elements, one switched in series ($R/\alpha^2 T_r$) and one in parallel ($1/k$) to $L_{th}$.

**Section S5: Pictures of the experimental set-ups**

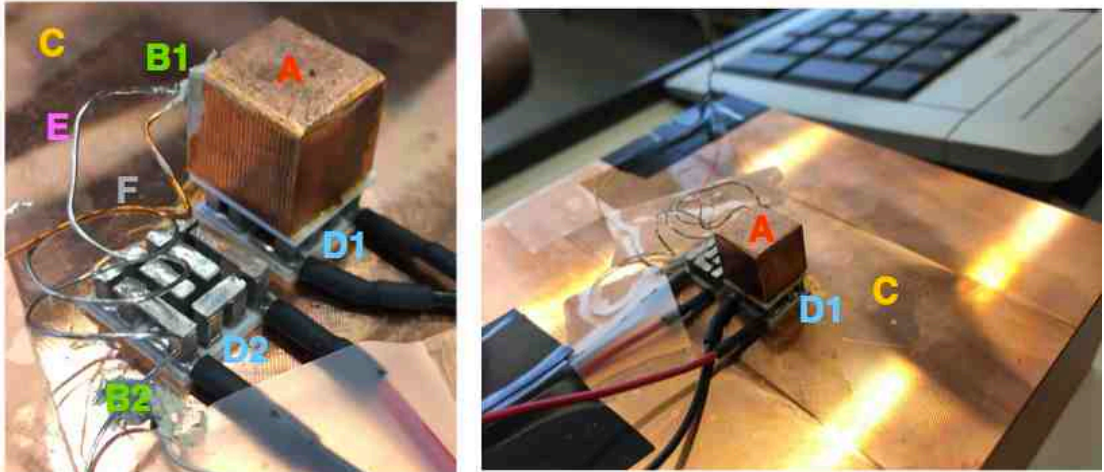

Copper cube as thermal load (A), placed on one side of the Peltier element (D1)

A    thermal load
B1   thermocouple leg at $T_b(t)$
B2   thermocouple leg at $T_r$
C    thermal bath
D1   Peltier element
D2   unused Peltier element (not connected)
E    constantan thermocouple wire
F    copper thermocouple wire

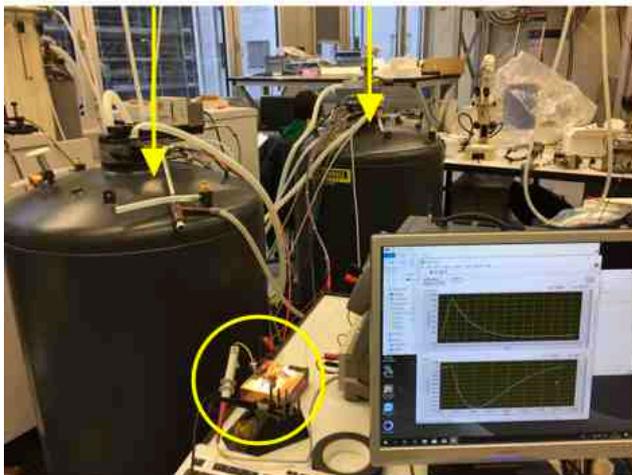

Experimental set-up of a configuration with two superconducting coils

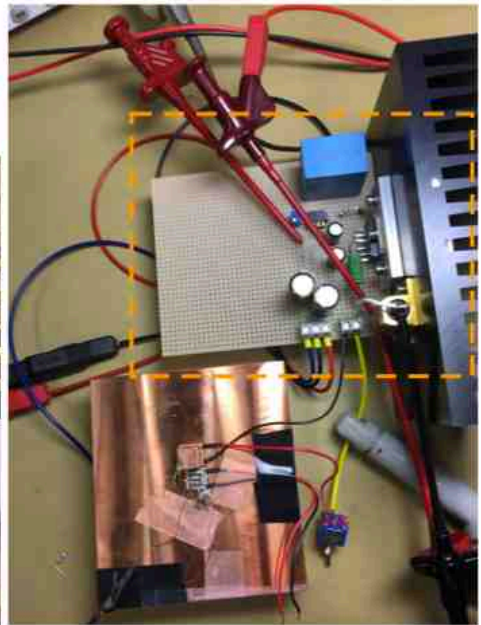

Experiment using a gyrator circuit

**Fig. S2. Experimental set-up of the oscillating thermal circuit**
Photo credit: A. Schilling

23